\documentclass[conference]{IEEEtran}

\usepackage[utf8]{inputenc}
\usepackage{amsmath}
\usepackage{amssymb}
\usepackage{graphicx}
\usepackage{booktabs}
\usepackage{multirow}
\usepackage{url}
\usepackage{hyperref}

\let\OLDthebibliography\thebibliography
\renewcommand\thebibliography[1]{
  \OLDthebibliography{#1}
  \footnotesize
  \setlength{\itemsep}{0pt}
  \setlength{\parskip}{0pt}
  \setlength{\parsep}{0pt}
}

\graphicspath{{./}{./models2/}}

\title{Decay-Region Group Delay as a Forensic Cue\\for AI-Generated Impulsive Sounds}

\author{
\IEEEauthorblockN{JaeHyeong Chang$^{*}$, Chengzhe Sun$^{*}$, and Siwei Lyu$^{\dagger}$}
\IEEEauthorblockA{
Institute for Artificial Intelligence and Data Science\\
University at Buffalo, Buffalo, NY, USA\\
\{jchang46, csun22, siweilyu\}@buffalo.edu\\
{\small $^{*}$Equal contribution.~~$^{\dagger}$Corresponding author.}
}
}

\begin{document}

\maketitle

\begin{abstract}
We investigate whether AI-generated impulsive sounds can be distinguished from real ones through group delay analysis. Our central finding is that AI-generated impulsive sounds show near-identical onset-region group-delay distributions but exhibit measurably different group-delay behavior in the late decay region: decay-region KL divergence reaches $0.322$ compared to near-zero onset divergence ($0.022$). Cross-band GD variability achieves single-feature AUC~=~0.720, and a Random Forest (RF) over nine decay-region features reaches AUC~$=$~0.884 under sample-disjoint evaluation. A group delay map used as a standalone 2D input to CNN classifiers achieves 90--94\% accuracy, demonstrating that group delay carries substantial discriminative information. Under generator hold-out, CNN and transformer classifiers show highly variable AUC (0.457--0.918). The group delay RF achieves the highest average hold-out accuracy among the evaluated methods ($66.7\%$) and avoids extreme below-random collapse, although its average AUC (0.731) is lower than CNN avg (0.762) and AST (0.772). Parameter sensitivity analysis across 27 STFT configurations confirms that the RF AUC remains stable (0.700--0.847, std~=~0.035). These results suggest that decay-region group delay can serve as a physically interpretable forensic cue that complements magnitude-based classifiers, while broader validation remains necessary.
\end{abstract}


\section{Introduction}

Generative audio models such as ElevenLabs, Stable Audio~\cite{evans2024stableaudio}, and AudioLDM2~\cite{liu2024audioldm2} can now synthesize impulsive sounds with increasing realism, raising concerns for forensics and misinformation detection~\cite{todisco2019asvspoof, yi2022survey}. Unlike speech deepfake detection, AI-generated environmental sound forensics remains comparatively underexplored. We study this problem through the lens of \emph{group delay} — the negative frequency-derivative of the phase spectrum — which characterizes how different frequency components are temporally delayed by an acoustic system. Real impulsive sounds are physically constrained: shock wave propagation, environmental reflection, and energy dissipation impose structured group delay patterns especially in the late decay region. We hypothesize that generative models, optimized for perceptual plausibility rather than physical consistency, may not fully reproduce this behavior.

The primary contribution of this paper is not a new state-of-the-art detector, but rather a demonstration that \textbf{decay-region group delay is an important forensic cue} for AI-generated impulsive sound detection. We make four contributions. First, we identify decay-region group delay inconsistency as a forensic cue observed across the three tested generators, supported by KL divergence, effect size, and classifier experiments. Second, group delay maps achieve 90--94\% standalone detection accuracy in the sample-disjoint setting. Third, we show that CNN and transformer classifiers exhibit highly variable AUC under generator hold-out (0.457--0.918), while the group delay RF avoids extreme below-random collapse and achieves higher average accuracy, suggesting that physically grounded features may provide complementary forensic cues. Fourth, parameter sensitivity analysis across 27 STFT configurations confirms robustness of the RF classifier (AUC std~$=$~0.035).

\begin{figure}[!t]
    \centering
    \includegraphics[width=0.94\columnwidth]{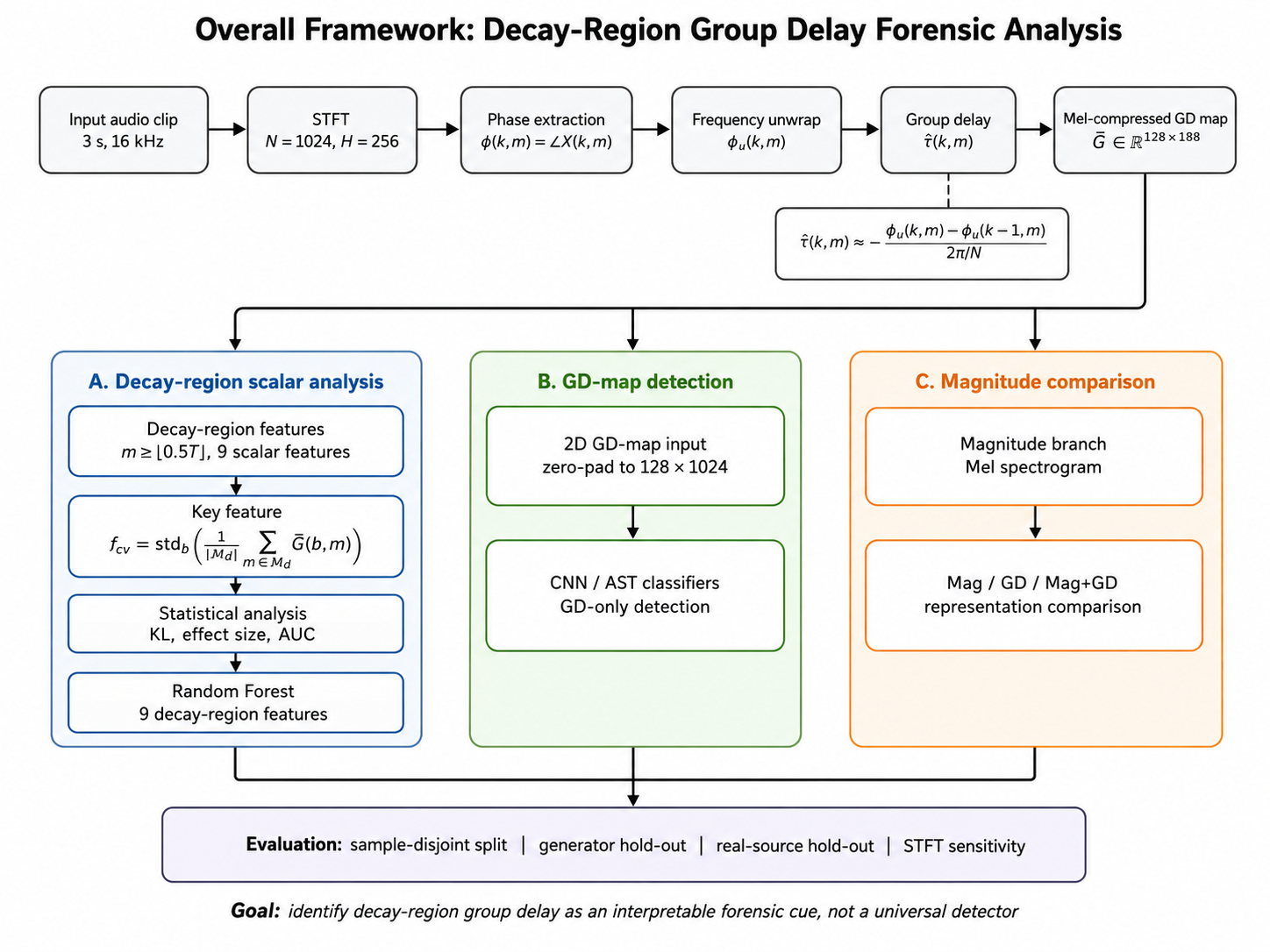}
    \caption{Overall framework for decay-region group delay forensic analysis. The pipeline processes a 3-second audio clip through STFT and phase unwrapping to extract a mel-compressed group delay map $\bar{G}$, which feeds three analysis branches: (A) decay-region scalar features with Random Forest classification, (B) GD-map-based CNN/AST detection, and (C) magnitude comparison. All branches are evaluated under sample-disjoint split, generator hold-out, real-source hold-out, and STFT sensitivity protocols.}
    \label{fig:framework}
\end{figure}

\section{Related Work}

\subsection{Audio Deepfake Detection}
The detection of AI-generated speech has been extensively studied through the ASVspoof challenge series~\cite{todisco2019asvspoof}, which has driven development of countermeasures based on spectral, cepstral, and phase features. Recent surveys~\cite{yi2022survey} highlight that magnitude-based classifiers achieve high in-distribution accuracy but often fail to generalize across unseen spoofing systems. Our work extends this line of research to the less-studied domain of impulsive sounds, where physical constraints on signal decay impose additional forensic structure.

\subsection{Phase and Group Delay in Audio Analysis}
The importance of phase information in signal reconstruction and analysis has long been recognized~\cite{oppenheim1981phase}. Group delay — the negative frequency-derivative of the phase spectrum — has been applied to speaker gender identification~\cite{lee2008gdf} and speech processing~\cite{yegnanarayana1984gdf}, where it captures temporal structure not visible in magnitude spectrograms. Phase-based features including group delay have been explored for speech deepfake detection~\cite{xue2022f0}, demonstrating that phase information complements magnitude features. Our work extends this to impulsive sound forensics, focusing specifically on decay-region group delay inconsistencies rather than full-spectrum phase features. To the best of our knowledge, decay-region group delay has not been systematically studied as a forensic cue for AI-generated impulsive sounds.

\subsection{Generative Audio Models}
Recent text-to-audio and sound synthesis models such as AudioLDM2~\cite{liu2024audioldm2} and Stable Audio~\cite{evans2024stableaudio} can generate perceptually convincing impulsive sounds. These models are optimized for perceptual quality rather than physical accuracy, motivating our hypothesis that they may fail to reproduce physically constrained signal properties such as decay-region group delay structure.

\section{Dataset}

We construct 15,000 clips (7,500 real / 7,500 fake) at 16\,kHz, 3 seconds, focusing on explosion and impact-type impulsive sounds. Table~\ref{tab:dataset} details sources and augmentation ratios. Real samples are drawn from FSD50K~\cite{fonseca2022fsd50k} and SESA~\cite{sesa2019}; fake samples from ElevenLabs, Stable Audio, and AudioLDM2. Augmentation applies time-shift, SpecAugment masking~\cite{park2019specaugment}, and noise mixing identically to both classes.

A \textbf{sample-disjoint} split ensures augmented variants of the same source file remain in the same partition, preventing augmentation leakage across Train/Val/Test~$=$~12000/1500/1500 (balanced). We acknowledge that the high augmentation ratio for SESA (6.4$\times$) and the source mismatch between real (FSD50K/SESA field recordings) and fake (synthesized audio) may introduce domain-level biases beyond real/fake distinction; we address this in the Limitations.

Because large-scale, controlled real impulsive sound datasets are difficult to obtain publicly, our dataset necessarily combines available real impulsive sound sources with generated samples. We therefore frame this work not as a definitive benchmark for universal impulsive sound deepfake detection, but as an initial forensic study showing that decay-region group delay contains discriminative and physically interpretable cues. Accordingly, results should be interpreted as evidence for a forensic cue rather than corpus-independent benchmark performance.

\begin{table}[!ht]
\centering
\caption{Dataset construction.}
\label{tab:dataset}
\begin{tabular}{llrrr}
\toprule
\textbf{Class} & \textbf{Source} & \textbf{Orig.} & \textbf{Final} & \textbf{$\times$} \\
\midrule
Real & FSD50K       & 1,122 & 3,750 & 3.3 \\
Real & SESA          &   585 & 3,750 & 6.4 \\
Fake & ElevenLabs   & 1,119 & 2,500 & 2.2 \\
Fake & Stable Audio & 1,000 & 2,500 & 2.5 \\
Fake & AudioLDM2    & 1,000 & 2,500 & 2.5 \\
\bottomrule
\end{tabular}
\end{table}

\subsection{Real-Source Bias Control}

To verify that our results are not driven by source-specific recording artifacts in the real set, we conducted a real-source hold-out experiment. In the FSD50K$\to$SESA condition, the GD RF is trained using FSD50K real samples and a sample-disjoint subset of fake samples, and tested using SESA real samples and held-out fake samples. The reverse condition (SESA$\to$FSD50K) swaps only the real source while preserving sample-disjoint fake partitions. Table~\ref{tab:source_bias} reports GD RF performance under each condition.

\begin{table}[!t]
\centering
\caption{Real-source hold-out: GD RF performance under real-source shift.}
\label{tab:source_bias}
\begin{tabular}{lcc}
\toprule
\textbf{Condition} & \textbf{AUC} & \textbf{Acc (\%)} \\
\midrule
FSD50K $\to$ SESA (hold-out)    & 0.849 & 76.47 \\
SESA $\to$ FSD50K (hold-out)    & 0.857 & 77.00 \\
Mixed $\to$ Mixed (reference)   & \textbf{0.878} & \textbf{79.33} \\
\bottomrule
\end{tabular}
\end{table}

The GD RF achieves AUC~=~0.849--0.857 under real-source shift vs.\ AUC~=~0.878 for the mixed reference. A sanity check classifying FSD50K vs.\ SESA real samples yields AUC~=~0.737, confirming partial source separability; results should therefore be interpreted as initial evidence for a forensic cue rather than a source-independent detector.

\section{Group Delay Representation}

\subsection{Group Delay Definition}

Group delay is formally defined as the negative derivative of the phase spectrum with respect to angular frequency~\cite{yegnanarayana1984gdf, lee2008gdf}:
\begin{equation}
    \tau(\omega) = -\frac{d\phi(\omega)}{d\omega}.
    \label{eq:gd_def}
\end{equation}
This quantity characterizes the time delay experienced by each frequency component as it passes through a system. In real impulsive sounds, $\tau(\omega)$ in the decay region reflects structured acoustic propagation constraints imposed by physical wave propagation and environmental response. The phase spectrum carries complementary information to magnitude~\cite{oppenheim1981phase}, and we approximate Eq.~(\ref{eq:gd_def}) via finite differences of the unwrapped STFT phase~\cite{oppenheim1999dsp}, as described below.

\subsection{STFT and Phase}

Given a 3-second waveform $x[n]$ at 16\,kHz, the Short-Time Fourier Transform is:
\begin{equation}
    X(k,m) = \sum_{n=0}^{N-1} x[n+mH]\,w[n]\,e^{-j2\pi kn/N},
\end{equation}
where $N=1024$, hop $H=256$, and $w[n]$ is a Hann window. The instantaneous phase is:
\begin{equation}
    \phi(k,m) = \angle\, X(k,m).
\end{equation}

\subsection{Group Delay Approximation}
\label{sec:approx}

We discretize Eq.~(\ref{eq:gd_def}) via finite differences of the frequency-unwrapped phase $\phi_u$:
\begin{equation}
    \hat{\tau}(k,m) \approx -\frac{\phi_u(k,m) - \phi_u(k-1,m)}{2\pi/N}.
    \label{eq:gd_approx}
\end{equation}
Values are clipped at $\tau_{\max} = 500$ samples ($\approx$31\,ms at 16\,kHz) for numerical stability:
\begin{equation}
    \tilde{\tau}(k,m) = \operatorname{clip}\bigl(\hat{\tau}(k,m),\,-\tau_{\max},\,\tau_{\max}\bigr).
\end{equation}

\subsection{Mel-compressed Group Delay Map}

As a practical representation, a mel filterbank $\{H_b(k)\}_{b=1}^{128}$ compresses the frequency axis:
\begin{equation}
    G(b,m) = \sum_{k} H_b(k)\,\tilde{\tau}(k,m),
\end{equation}
normalized to $[-1,1]$:
\begin{equation}
    \bar{G}(b,m) = \frac{G(b,m)}{\max_{b,m}|G(b,m)| + \epsilon}.
    \label{eq:gdmap}
\end{equation}
For a 3-second clip at 16\,kHz with hop $H=256$, this yields $\bar{G} \in \mathbb{R}^{128 \times 188}$, zero-padded to $128 \times 1024$ only to match the input resolution expected by image-based classifiers; scalar GD features are computed from the original $128 \times 188$ representation. The same temporal zero-padding procedure was applied consistently across all compared image-based representations (Mag, GD, Mag+GD) to avoid representation-specific input-size differences.

\subsection{Decay-Region Features}

We define the late temporal region as frames $m \geq \lfloor 0.5T \rfloor$, where $T=188$ is the total number of frames, and refer to it as the decay region, as impulsive energy predominantly occurs in the earlier portion of the three-second clips. The most discriminative scalar feature is \textbf{cross-band GD variability}:
\begin{equation}
    f_{\text{cv}} = \operatorname{std}_b\!\left(\frac{1}{|\mathcal{M}_d|}\sum_{m\in\mathcal{M}_d} \bar{G}(b,m)\right),
    \label{eq:fc}
\end{equation}
measuring the standard deviation of per-band mean group delay in the decay region. AI-generated samples show \emph{higher} cross-band GD variability than real impulsive sounds, suggesting less stable or less physically consistent phase-derivative behavior across frequency bands. Real impulsive sounds show \emph{lower} and more structured variability, consistent with physically constrained decay. The \textbf{Decay-to-Onset KL divergence} quantifies temporal distributional shift:
\begin{equation}
    f_{\text{KL}} = D_{\mathrm{KL}}\!\left(p_{\text{decay}} \,\|\, p_{\text{onset}}\right).
\end{equation}

\section{Group Delay Forensic Analysis}

\subsection{Global vs.\ Decay-Region Separability}

\begin{figure}[!t]
    \centering
    \includegraphics[width=0.94\columnwidth]{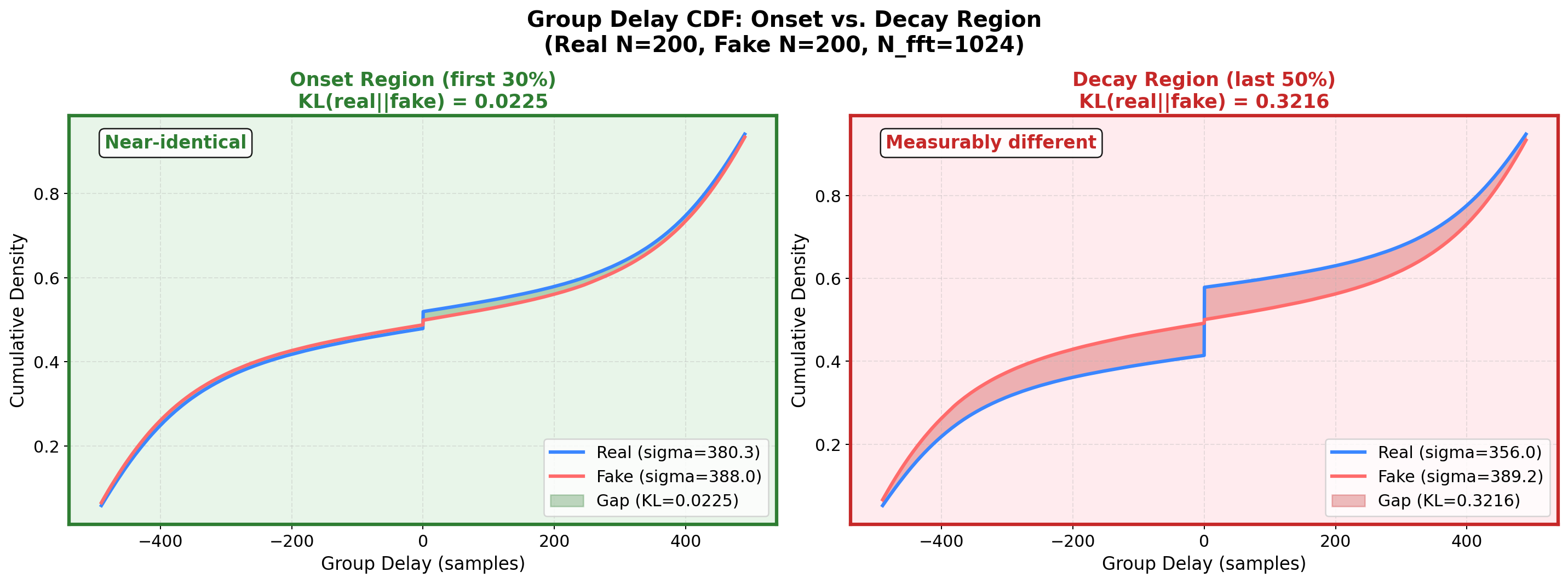}
    \caption{Group delay CDF: onset region (left, KL~=~0.022) shows near-identical real/fake distributions, while the decay region (right, KL~=~0.322) reveals measurably different distributions. Shaded area indicates the gap between real and fake CDFs.}
    \label{fig:gd_histogram}
\end{figure}

Figure~\ref{fig:gd_histogram} shows onset and decay region CDFs; Table~\ref{tab:kl} additionally reports global and frequency-band KL values, all of which are approximately 0.24, indicating limited global separability. As shown in Fig.~\ref{fig:gd_histogram} (left panel), the onset region shows near-identical distributions (KL~=~0.022), indicating that onset GD alone does not distinguish the two classes. The decay region (right panel) reveals measurably larger divergence (KL~=~0.322), localizing forensic artifacts in late decay.

\begin{table}[!t]
\centering
\caption{KL divergence: real vs.\ fake.}
\label{tab:kl}
\begin{tabular}{lc}
\toprule
\textbf{Region / Band} & \textbf{KL} \\
\midrule
Overall & 0.2426 \\
Low freq (0--500\,Hz) & 0.2339 \\
Mid freq (500--2000\,Hz) & 0.2417 \\
High freq (2000--8000\,Hz) & 0.2437 \\
Onset (first 30\%)              & 0.022 \\
\textbf{Decay (last 50\%)}      & \textbf{0.322} \\
\bottomrule
\end{tabular}
\end{table}

\subsection{Scalar Feature Analysis}

Table~\ref{tab:features} reports effect sizes and AUC ($N=15{,}000$, 5-fold CV, all $p<0.001$). Cross-band GD variability ($f_{\text{cv}}$) achieves the highest discriminative power (Cohen's $d=0.646$, AUC~=~0.720): AI-generated samples show higher cross-band GD variability than real impulsive sounds, suggesting less stable or less physically consistent phase-derivative behavior. Real impulsive sounds show lower and more structured variability. The Decay-to-Onset KL divergence shows striking distributional separation (real: $\mu=0.923$, $\sigma=1.795$; all generators: $\mu\approx0.010$, $\sigma=0.010$). Its standalone AUC of 0.570, however, reflects the high variance in the real set ($\sigma=1.795$): while the mean difference is large, the real distribution is highly spread, reducing class separability at the individual sample level. The Decay-to-Onset KL should therefore be interpreted as a \textbf{distribution-level forensic cue} rather than a strong standalone sample-level classifier. A Random Forest~\cite{breiman2001rf} over all nine features achieves AUC~=~0.884 under sample-disjoint evaluation (all 15,000 samples; note this differs from the mixed real-source reference AUC~=~0.878 in Table~\ref{tab:source_bias}, which uses an 80/20 train/test split on a subset). Note that the KL values in Fig.~\ref{fig:gd_histogram} measure distributional divergence between real and fake samples within each temporal region, whereas the Decay-to-Onset KL feature ($f_{\text{KL}}$) in Table~\ref{tab:features} measures within-sample temporal shift between onset and decay distributions.

\begin{table}[!t]
\centering
\caption{Decay-region scalar features ($N=15{,}000$, all $p<0.001$).}
\label{tab:features}
\begin{tabular}{lccc}
\toprule
\textbf{Feature} & \textbf{Cohen's $d$} & \textbf{Cliff's $\delta$} & \textbf{AUC} \\
\midrule
Decay variance          & 0.588 & 0.261 & 0.651 \\
Decay entropy           & 0.502 & 0.010 & 0.521 \\
Decay abs.\ mean        & 0.578 & 0.204 & 0.628 \\
Decay/onset var.\ ratio & 0.036 & 0.096 & 0.573 \\
Decay$\to$onset KL      & 0.482 & 0.114 & 0.570 \\
Decay$\to$onset shift   & 0.084 & 0.102 & 0.530 \\
\textbf{Cross-band GD var.\ ($f_{\text{cv}}$)} & \textbf{0.646} & \textbf{0.424} & \textbf{0.720} \\
Temporal fluct.         & 0.531 & 0.054 & 0.566 \\
Global variance         & 0.604 & 0.306 & 0.675 \\
\bottomrule
\end{tabular}
\end{table}

\subsection{Consistency Across Generators}

\begin{figure}[!t]
    \centering
    \includegraphics[width=0.94\columnwidth]{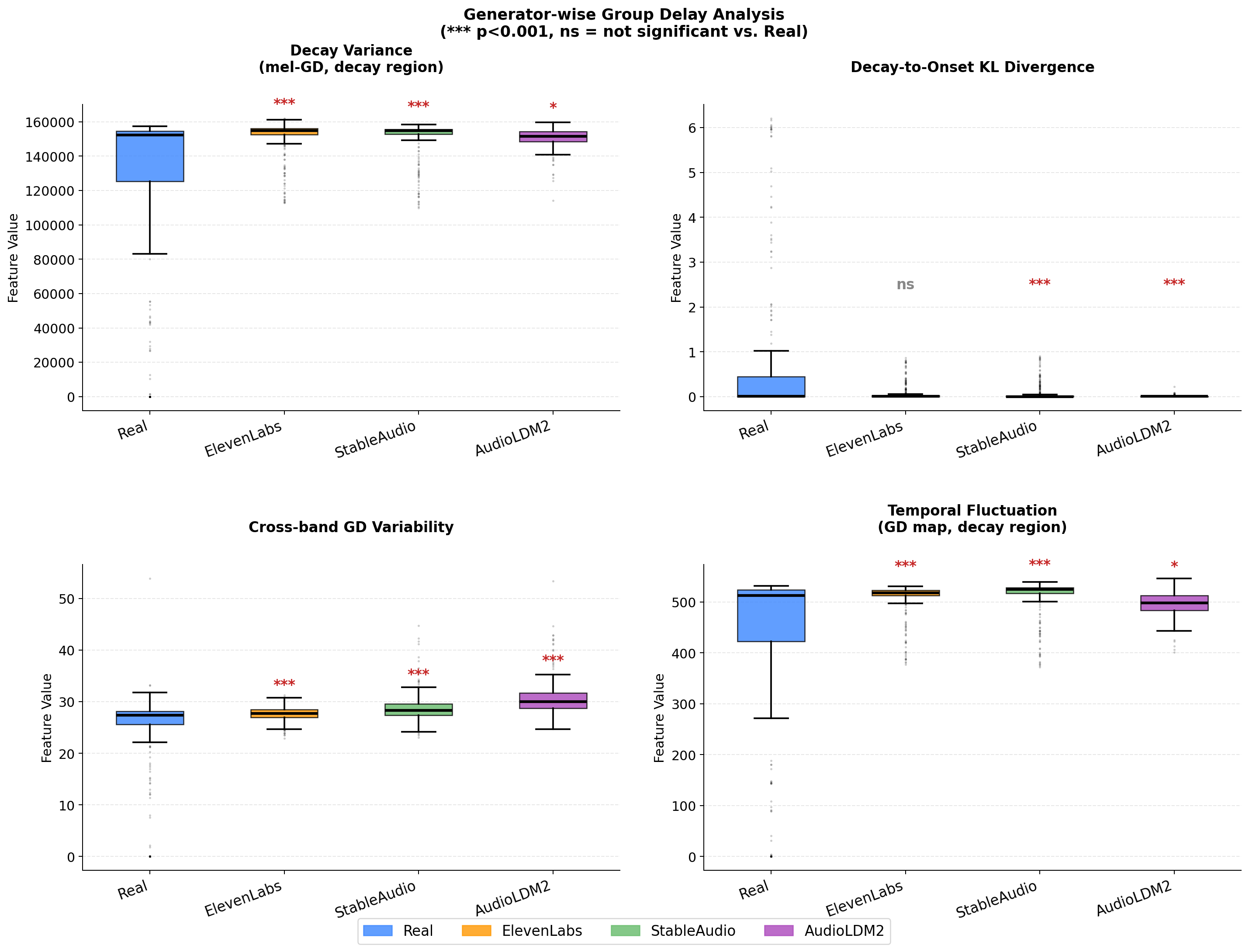}
    \caption{Generator-wise group delay analysis ($^{***}p<0.001$, $^{*}p<0.05$, ns~=~not significant vs.\ real). The most pronounced separation appears in Decay-to-Onset KL (top-right), where Stable Audio and AudioLDM2 differ significantly ($p<0.001$) while ElevenLabs shows a weaker, non-significant trend. The compact fake distributions may partly reflect generator-specific homogeneity; these results are interpreted as evidence of a candidate forensic cue rather than a source-independent detector.}
    \label{fig:generator}
\end{figure}

Figure~\ref{fig:generator} shows group delay feature distributions across the three tested generators. The most visually pronounced separation appears in the Decay-to-Onset KL feature (top-right), where Stable Audio and AudioLDM2 cluster near zero while real samples show large variability ($\mu=0.923$, $p<0.001$); ElevenLabs shows a weaker but directionally consistent trend (not significant on this feature alone). Cross-band GD variability and decay variance show statistically significant but partially overlapping differences across the tested generators, with varying significance levels. These results suggest a systematic tendency in decay-region group delay behavior of the tested generators, though the pattern is not uniformly strong across all features and generators.

\subsection{Parameter Sensitivity Analysis}

We evaluated group delay features across 27 STFT configurations: $N \in \{512, 1024, 2048\}$, hop $\in \{128, 256, 512\}$, clip $\in \{250, 500, 1000\}$ ($N_{\text{sample}}=500$ per class, 5-fold CV). Table~\ref{tab:param} reports results averaged over clip values for each $(N, \text{hop})$ pair.

\begin{table}[!t]
\centering
\caption{STFT parameter sensitivity (27 configs, avg.\ over clip values).}
\label{tab:param}
\begin{tabular}{rrcc}
\toprule
\textbf{$N_{\text{fft}}$} & \textbf{hop} & \textbf{$f_{\text{cv}}$ AUC} & \textbf{RF AUC} \\
\midrule
512  & 128 & 0.710 & 0.801 \\
512  & 256 & 0.557 & 0.755 \\
512  & 512 & 0.569 & 0.768 \\
\midrule
1024 & 128 & 0.730 & 0.841 \\
1024 & 256$^{*}$ & 0.687 & 0.828 \\
1024 & 512 & 0.584 & 0.752 \\
\midrule
2048 & 128 & 0.728 & 0.822 \\
2048 & 256 & 0.725 & 0.804 \\
2048 & 512 & 0.696 & 0.777 \\
\bottomrule
\end{tabular}
\vspace{2pt}

{\footnotesize $^{*}$Default configuration used in the main experiments.}
\end{table}

The RF AUC ranges 0.700--0.847 (std~=~0.035) across all 27 configurations. Cross-band GD variability shows higher sensitivity (std~=~0.069), with lower performance for $N=512$; for $N \geq 1024$ the AUC stabilizes at $0.686$--$0.734$. The default ($N=1024$, hop~=~256) is representative of the stable region.

\section{Detection Experiments}

\subsection{Feature Representation Comparison}

We trained ResNet50~\cite{he2016deep}, EfficientNet-B2~\cite{tan2019efficientnet}, CNN14~\cite{kong2020panns}, and AST~\cite{gong2021ast, dosovitskiy2021vit} with three inputs for 15 epochs under the sample-disjoint protocol: \textbf{Mag} (mel-spectrogram, 1-ch), \textbf{GD} (group delay map $\bar{G}$, 1-ch), \textbf{Mag+GD} (2-ch). Results are shown in Table~\ref{tab:repr}.

\begin{table}[!t]
\centering
\caption{Feature representation (sample-disjoint, 15 epochs, accuracy \%).}
\label{tab:repr}
\begin{tabular}{lrrr}
\toprule
\textbf{Model} & \textbf{Mag} & \textbf{GD only} & \textbf{Mag+GD} \\
\midrule
ResNet50        & 97.00 & 90.40 & \textbf{97.40} \\
EfficientNet-B2 & \textbf{98.93} & 92.67 & 98.33 \\
CNN14           & \textbf{98.93} & 94.20 & 98.73 \\
AST             & \textbf{98.27} & 93.67 & 97.60 \\
\bottomrule
\end{tabular}
\end{table}

Models were trained with AdamW~\cite{loshchilov2019adamw}. \textbf{GD-only achieves 90--94\%} standalone accuracy, demonstrating substantial discriminative information in group delay maps. Mag+GD improves ResNet50 by 0.4 pp; higher-capacity models show a ceiling effect. The gap between in-distribution and hold-out performance suggests that only part of GD's discriminative information transfers across unseen generators.

\subsection{Generator Hold-out Evaluation}

Fake samples from one generator were excluded entirely from training and used only for testing. Each test set was balanced between real and the held-out fake generator. Table~\ref{tab:holdout} reports accuracy and AUC for all evaluated models.

\begin{table}[!t]
\centering
\caption{Generator hold-out: Acc (\%) and AUC. \textbf{Bold}: best Acc or best AUC per hold-out condition. CNN avg denotes the average over ResNet50, EfficientNet-B2, and CNN14; AST is reported separately as a transformer-based model.}
\label{tab:holdout}
\begin{tabular}{llcc}
\toprule
\textbf{Hold-out} & \textbf{Method} & \textbf{Acc} & \textbf{AUC} \\
\midrule
\multirow{5}{*}{AudioLDM2}
 & ResNet50        & \textbf{61.2} & \textbf{0.714} \\
 & EfficientNet-B2 & 54.7 & 0.586 \\
 & CNN14           & 55.2 & 0.457 \\
 & AST             & 57.2 & 0.659 \\
 & GD RF           & 58.6 & 0.580 \\
\midrule
\multirow{5}{*}{Stable Audio}
 & ResNet50        & 64.1 & 0.829 \\
 & EfficientNet-B2 & 63.1 & 0.907 \\
 & CNN14           & 69.0 & \textbf{0.918} \\
 & AST             & 69.1 & 0.906 \\
 & \textbf{GD RF}  & \textbf{73.6} & 0.832 \\
\midrule
\multirow{5}{*}{ElevenLabs}
 & ResNet50        & 59.1 & \textbf{0.825} \\
 & EfficientNet-B2 & 54.3 & 0.820 \\
 & CNN14           & 54.1 & 0.806 \\
 & AST             & 53.6 & 0.753 \\
 & \textbf{GD RF}  & \textbf{67.8} & 0.782 \\
\midrule
\multirow{3}{*}{Average}
 & CNN avg (ResNet/EffNet/CNN14) & 59.4 & 0.762 \\
 & AST             & 59.9 & \textbf{0.772} \\
 & \textbf{GD RF}  & \textbf{66.7} & 0.731 \\
\bottomrule
\end{tabular}
\end{table}

CNN and transformer (AST) classifiers show \textbf{highly variable AUC} (0.457--0.918): CNN14 reaches AUC~=~0.918 for Stable Audio hold-out but collapses to AUC~=~0.457 for AudioLDM2 — below random. AST similarly ranges from 0.659 to 0.906, suggesting that both CNN and transformer architectures may rely substantially on generator-specific artifacts rather than fully generator-generalizable real/fake cues. The GD RF does not achieve the highest average AUC (0.731 vs.\ 0.762 for CNN avg and 0.772 for AST), but it avoids the extreme below-random collapse observed for CNN14 (AUC~=~0.457) and yields the highest average accuracy (66.7\% vs.\ 59.4\% for CNN avg and 59.9\% for AST).

The AudioLDM2 hold-out is the most challenging condition for all methods; the GD RF also struggles here (AUC~=~0.580), suggesting that group delay cues do not transfer uniformly across generator architectures. These results indicate that group delay features provide complementary forensic cues, though further validation is needed.

\subsection{Supplementary: Phase-Flow CRNN Attention}

\begin{figure}[!t]
    \centering
    \includegraphics[width=0.94\columnwidth]{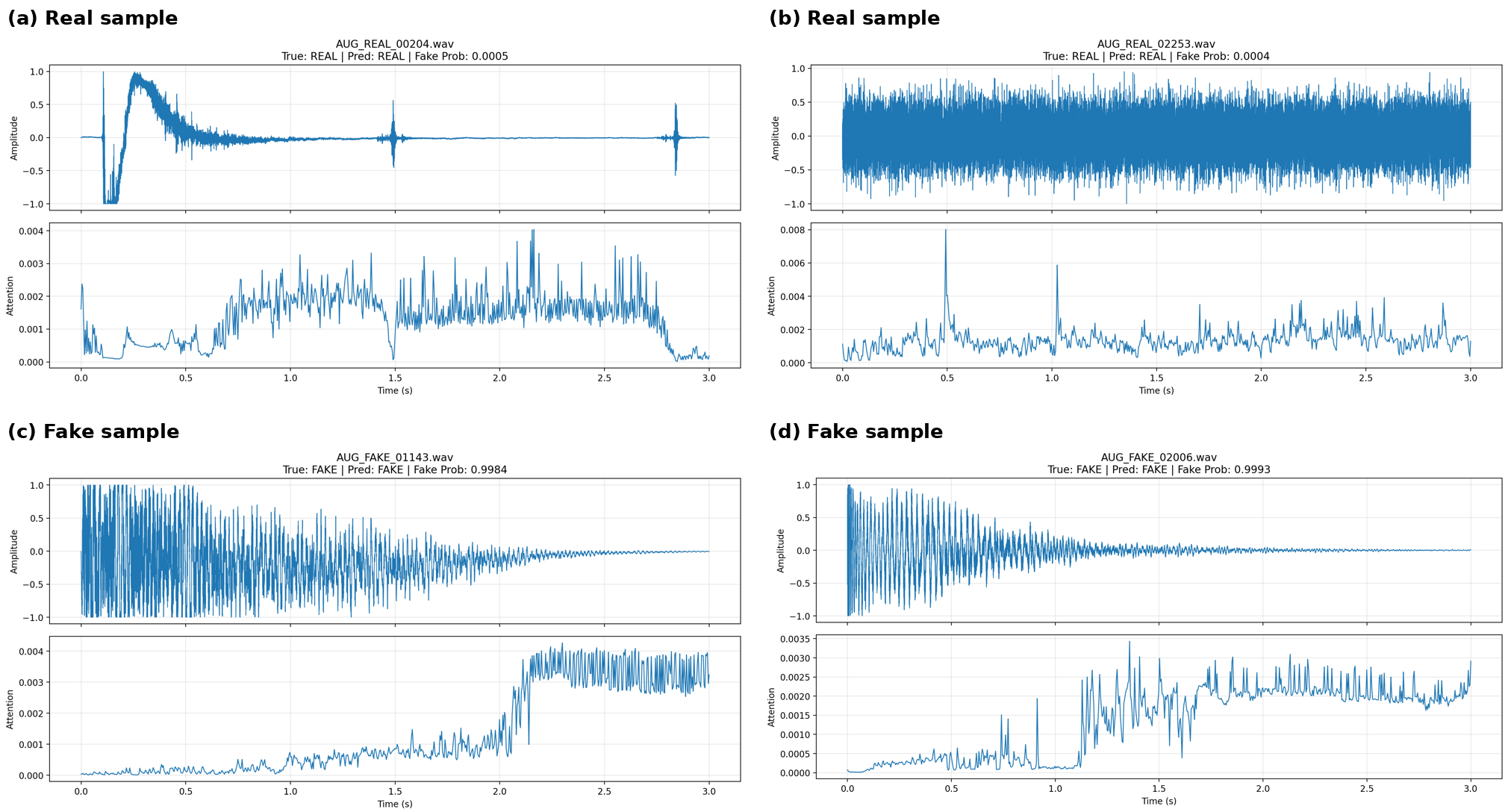}
    \caption{Phase-Flow CRNN temporal attention (note: y-axis scales differ across panels). Real samples (a,b) show relatively diffuse or transient attention patterns, whereas fake samples (c,d) exhibit concentrated attention in the late decay region. This pattern is consistent with the decay-region group delay findings.}
    \label{fig:attention}
\end{figure}

A lightweight Phase-Flow CRNN~\cite{choi2017crnn} (2.24M parameters) processes temporal phase-flow via BiGRU~\cite{cho2014gru} with temporal attention, achieving 92.33\% in-distribution accuracy. Its attention concentrates on late decay regions (Figure~\ref{fig:attention}), providing qualitative support for the decay-region group delay finding.

\section{Conclusion}

Real impulsive sounds undergo physically constrained group delay evolution from onset to decay (Eq.~\ref{eq:gd_def}). The tested generators show substantially lower within-sample Decay-to-Onset KL values than real impulsive sounds, which exhibit greater variability. GD-only maps achieve 90--94\% standalone accuracy in the sample-disjoint setting. Under generator hold-out, CNN and transformer classifiers show highly variable AUC (0.457--0.918), while the group delay RF achieves the highest average accuracy and avoids extreme below-random collapse, despite not achieving the highest average AUC. Parameter sensitivity analysis confirms RF robustness across 27 STFT configurations (AUC std~$=$~0.035). Phase-Flow CRNN attention provides qualitative support for the decay-region locus. These results show that \emph{high in-distribution accuracy does not imply generator-generalizable forensic understanding}. Phase-based representations — specifically decay-region group delay — offer a physically interpretable and complementary lens for AI-generated impulsive sound forensics.

\section{Limitations and Future Work}

Several limitations should be acknowledged. Real data relies on augmentation (SESA: 6.4$\times$), and source/domain mismatch may introduce biases (FSD50K vs.\ SESA GD AUC~=~0.737), though the real-source hold-out experiment in Section III-A shows this bias is modest. Only three generators are evaluated, and generalization is not yet uniform: the GD RF collapses to near-chance performance on AudioLDM2 hold-out (AUC~=~0.580), while remaining more stable for Stable Audio and ElevenLabs. This pattern, together with the Decay-to-Onset KL feature's gap between strong distribution-level separation (real $\mu=0.923$ vs.\ generators $\mu\approx0.010$) and weaker sample-level discriminability (AUC~=~0.570), suggests decay-region group delay is best understood as a distribution-level forensic signature rather than a universal sample-level classifier -- a distinction we believe is itself a useful finding for future detector design. The STFT parameter sensitivity analysis, based on a smaller subset (500 samples per class), and the absence of adversarial robustness testing (e.g., against phase randomization) are natural next steps, alongside extending evaluation to sound categories beyond explosions and impacts and to clip durations other than 3 seconds. $N_{\text{fft}} \geq 1024$ is recommended based on current results. We view these as concrete, addressable directions rather than fundamental obstacles to the core finding that decay-region group delay carries forensically meaningful signal.

\newpage
\bibliographystyle{IEEEtran}
\bibliography{references}

\end{document}